\documentclass{ptephy_v1}

\usepackage{amsmath} 
\usepackage{amsthm} 
\usepackage{amssymb} 
\usepackage{latexsym}
\usepackage{graphicx}

\newcommand{\eqb}{\begin{equation}}
\newcommand{\eqe}{\end{equation}}
\newcommand{\eqbnon}{\begin{equation*}}
\newcommand{\eqenon}{\end{equation*}}

\newcommand{\eqab}{\begin{eqnarray}}
\newcommand{\eqae}{\end{eqnarray}}
\newcommand{\eqabnon}{\begin{eqnarray*}}
\newcommand{\eqaenon}{\end{eqnarray*}}

\newcommand{\seqb}{\begin{subequations}} 
\newcommand{\seqe}{\end{subequations}}   

\newcommand{\eref}[1]{\eqref{#1}} 
\newcommand{\Fref}[1]{Figure~\ref{#1}}
\newcommand{\fref}[1]{figure~\ref{#1}}

\newcommand{\defeq}{:=}

\newcommand{\od}[2]{\frac{{\rm d} #1}{{\rm d} #2}}
\newcommand{\diff}{{\rm d}}

\newcommand{\ls}[1]{_{\rm #1}}

\newcommand{\bound}{\mathcal{U}}
\newcommand{\ncor}{n\ls{cor}}
\newcommand{\nenv}{n\ls{env}}
\newcommand{\Kenv}{K\ls{env}}

\begin{document}

\title{Spherical polytropic balls cannot mimic black holes}

\author{
\name{Hiromi Saida}{1}, 
\name{Atsuhito Fujisawa}{2}, 
\name{Chul-Moon Yoo}{2}, 
and \name{Yasusada Nambu}{2}
}


\address{
\affil{1}{Department of Physics, Daido University, Minami-ku, 457-8530 Nagoya, Japan}
\affil{2}{Department of Physics, Nagoya University, Chikusa-ku, 464-8602 Nagoya, Japan}
\email{saida@daido-it.ac.jp , fujisawa@gravity.phys.nagoya-u.ac.jp , yoo@gravity.phys.nagoya-u.ac.jp , nambu@gravity.phys.nagoya-u.ac.jp}
}

\begin{abstract}
The so-called black hole shadow is a dark region which is expected to appear in a fine image of optical observation of black holes. 
It is essentially an absorption cross section of black hole, and the boundary of shadow is determined by unstable circular orbits of photons (UCOP).
If there exists a compact object possessing UCOP but no black hole horizon, it can provide us with the same shadow image with black holes, and a detection of shadow image cannot be a direct evidence of black hole existence. 
Then, this paper examine whether or not such compact objects can exist under some suitable conditions.
We investigate thoroughly the static spherical polytropic ball of perfect fluid with single polytrope index, and then investigate a representative example of the piecewise polytropic ball. 
Our result is that the spherical polytropic ball which we have investigated cannot possess UCOP, if the sound speed at center is subluminal (slower-than-light). 
This means that, if the polytrope treated in this paper is a good model of stellar matter in compact objects, the detection of shadow image is regarded as a good evidence of black hole existence. 
As a by-product, we have found the upper bound of the mass-to-radius radio of polytropic ball with single index, $M_\ast/R_\ast < 0.281$, under the subluminal-sound-speed condition. 
\end{abstract}

\subjectindex{E31, E30, E01}

\maketitle

\section{Introduction and our aim}
\label{sec:intro}

In recent years, the resolution of image by the very-long-baseline-interferometer (VLBI) radio observation is approaching to the visible angular size of SgrA$^\ast$, about $10$~micro-arcsecond (a black hole candidate of $4\times 10^6\,M_{\odot}$ at the center of our galaxy, $8$~kpc from the earth), which is the largest visible angular size in known black hole candidates~\cite{ref:miyoshi+4.2004,ref:doeleman+etal.2008}. 
The so-called black hole shadow is expected to be resolved by such fine observation near future (see~\cite{ref:takahashi.2004,ref:kanai+1.2013} and references therein). 
It seems to be currently a common understanding that seeing the black hole shadow is believing the existence of black hole horizon.

However, this common understanding has not been confirmed in general relativity as follows: 
Remember that the black hole shadow is a dark region appearing in an optical image of black holes, on which some photons would be detected if the black hole did not exist. 
Therefore, the shadow is essentially an absorption cross section of black hole. 
However, it should be emphasized that photons on the edge of shadow have been circulating around black hole before coming to the observer. 
The innermost circular orbit of those photons is not a great circle on black hole horizon, but an unstable circular orbit of photons (UCOP).
That is to say, the boundary of shadow is determined not by the black hole horizon, but by the UCOP.
This indicates that the direct origin of shadow is the UCOP, not the black hole horizon. 
Hence, although we can conclude the existence of UCOP once a shadow is observed, however, we cannot conclude immediately the existence of black hole horizon even if a shadow is clearly observed.

Here, let us assign a term, black hole mimicker, to a compact object possessing UCOP but no black hole horizon. 
If there exists a black hole mimicker, it can provide us with the same shadow image with black holes in optical observation, and a detection of shadow image cannot be a direct evidence of black hole existence. 
Therefore, we are interested in an existence/non-existence condition of black hole mimickers.

Some exotic candidates of black hole mimicker have been proposed, such as gravastars and boson stars. 
Under the assumption that any black hole mimicker emits a thermal radiation from its surface, the current observational data of SgrA$^\ast$ (the mass accretion rate and the observed flux) exclude the possibility that SgrA$^\ast$ may be those exotic black hole mimickers~\cite{ref:broderick+1.2006}. 
Further, the current observational data of some black hole candidates indicate that the gravastar cannot exist in nature, if any gravastar emits always a thermal radiation from its surface~\cite{ref:broderick+1.2007}. 
For the case that the gravastar does not emit radiations from its surface, the general feature of gravastar shadow, which enables us to distinguish gravastars from black holes in the shadow image, has already been examined~\cite{ref:sakai+2.2014}. 
The investigation on those exotic models may be interesting. 
However, we focus on a rather usual model in this paper.

Consider a static spherical ball of perfect fluid matter in the framework of general relativity, which connects to Schwarzschild geometry at its surface. 
A fluid ball, which does not possess the black hole horizon, becomes a black hole mimicker if it possesses one of following properties:
\begin{description}
\item[(A) ]
The ball is so compact that there appears a UCOP in the outside Schwarzschild geometry. 
\item[(B) ]
The ball is not so compact as case (A), but a UCOP appears inside the ball. 
\end{description}
In case (A), if the surface of fluid ball neither emit nor reflect any radiation, this ball can provide us with the same shadow image with a black hole. 
In case (B), if the fluid outside UCOP is completely transparent and if the fluid inside UCOP is not transparent, this ball can provide us with the same shadow image with a black hole.

Concerning the case (A), the mass-to-radius radio $3M_\ast/R_\ast$ of fluid ball is the key quantity, where $M_\ast$ and $R_\ast$ are respectively the total mass and surface radius of the ball measured in the dimension of length. 
If this ratio is less than unity ($3M_\ast/R_\ast < 1$), then no UCOP appears outside the fluid ball because the radius of UCOP in Schwarzschild geometry is $3M_\ast$. 
Note that, in order to avoid the gravitational collapse, an inequality, $3M_\ast/R_\ast < 3/2$, must hold. 
Further, by adding some reasonable conditions to the fluid ball, the upper bound of the ratio, $3M_\ast/R_\ast \le \bound$, should decrease,
\eqb
 \frac{3M_\ast}{R_\ast} \le \bound < \frac{3}{2} \,.
\eqe
Hence, the problem in case (A) is whether or not the upper bound $\bound$ becomes less than unity under some reasonable conditions of the fluid ball.

The upper bound $\bound$ has already been estimated for some situations. 
For the fluid ball with any equation of states satisfying three conditions, non-increasing energy density in outward direction, barotropic form of equation of states, and subluminal (slower-than-light) sound speed, an upper bound has been obtained by our previous work~\cite{ref:fujisawa+3.2015}, $\bound \simeq 1.0909209$. 
And, for a core-envelope model of neutron star~\cite{ref:hartle.1978,ref:iyer+2.1985}, the other value of upper bound has been obtained, $\bound \simeq 1.018$, where the equation of states in the envelope region is determined by nuclear matter physics (see section~2 of~\cite{ref:iyer+2.1985} for details), while the equation of states in the core region is treaded by the same method used in our previous work~\cite{ref:fujisawa+3.2015}. 
These estimations give still a bound greater than unity, $\bound > 1$. 
Further investigation is necessary to consider what kind of equation of states results in $\bound < 1$.

On the other hand, concerning the case (B), there is no existing work analyzing UCOP inside a fluid ball as far as we know. 
We need to formulate the criterion for judging the existence/non-existence of UCOP, and then apply the criterion to our model of fluid ball.

In this paper, we regard the polytrope as a representative example of the barotropic equation of states. 
The model investigated here is the static spherical ball of perfect fluid with the polytropic equation of states. 
Our investigation focuses mainly on the simple polytrope model whose polytrope index is fixed at one value for all region inside the fluid ball. 
After the thorough investigation of the simple model, an example of the piecewise polytropic fluid ball~\cite{ref:muller+1.1985} is investigated in the frame work of core-envelope model, where the value of polytrope index in the core region differs from that in the envelope region. 
Although the thorough study of piecewise polytrope model is left for next tasks, we can find a good insight into the core-envelope model of polytropic fluid balls.

For both polytrope models, we examine whether or not the cases (A) and (B) are possible for the polytropic fluid ball. 
Our result is that there cannot exist UCOP in neither outside nor inside of the polytropic fluid balls, if the sound speed at center is subluminal. 
This implies that, if the polytrope investigated in this paper is a good model of stellar matter in compact objects, a detection of shadow image is regarded as a good evidence of existence of black hole.\footnote{
There have been some attempts to study a nonlinear instability of black hole mimickers (see~\cite{ref:cardoso+4.2014} and references therein). 
This is an interesting approach. 
However, in those works~\cite{ref:cardoso+4.2014}, while the argument for nonlinear instability of black hole mimicker is conjectured by combination of linear analyses, but no definite proof of nonlinear instability has been obtained. 
}

In section~\ref{sec:ball}, a set-up of our analysis is described. 
Sections~\ref{sec:a} and~\ref{sec:b} are devoted to analyses of, respectively, cases (A) and (B) for the simple polytrope model. 
In section~\ref{sec:ce}, the analyses in previous sections are extended to an example of the core-envelope piecewise polytrope model. 
Section~\ref{sec:sd} is for summary and discussions.

\section{Static spherical polytropic ball}
\label{sec:ball}

We consider static spherical ball made of polytropic perfect fluid. 
The metric of this spacetime is given by a line element,
\eqb
\label{eq:ball.metric}
 \diff s^2 = g_{\mu\nu}\diff x^{\mu}\, \diff x^{\nu} =
 - e^{2\Phi(r)}c^2 \diff t^2 + \frac{\diff r^2}{1-2Gm(r)/(c^2 r)}
 + r^2 ( \diff \theta^2 + \sin^2\theta \diff \varphi^2 ) \,,
\eqe
where $(t,r,\theta,\varphi)$ is spherical poler coordinates, $\Phi(r)$ gives the lapse function, and $m(r)$ is the mass of fluid contained in the spherical region of radius $r$. 
The energy-momentum tensor of perfect fluid is $T_{\mu\nu} = [\,\sigma(r)c^2 + p(r)\,] u_\mu u_\nu + p(r) g_{\mu\nu}$, where $u = e^{-\Phi}\partial_{ct}$ is the four-velocity of static fluid, and $\sigma(r)$ and $p(r)$ are respectively the mass density and pressure of fluid.

A condition, $m(0) = 0$, should hold due to the regularity of spacetime at center. 
This implies a finite mass density at center, $\sigma\ls{c} = \sigma(0) \neq \infty$, where the suffix c denotes the value at center. 
We normalize all quantities by $\sigma\ls{c}$,
\eqb
\label{eq:ball.normalization}
 R \defeq \frac{\sqrt{G \sigma\ls{c}}}{c}\,r
 \quad,\quad
 \Sigma(R) \defeq \frac{\sigma(r)}{\sigma\ls{c}}
 \quad,\quad
 M(R) \defeq \frac{\sqrt{G^3 \sigma\ls{c}}}{c^3}\,m(r)
 \quad,\quad
 P(R) \defeq \frac{p(r)}{\sigma\ls{c} c^2} \,.
\eqe
These are dimension-less. 
The lapse function, $\Phi(r) \defeq \Phi(R)$, does not need normalization because $\Phi$ is already dimension-less by definition~\eref{eq:ball.metric}.

The barotropic equation of states is generally expressed as $P = P(\Sigma)$. 
We adopt the polytrope as a representative form of the barotropic matter,
\seqb
\label{eq:ball.eos}
\eqb
 P(\Sigma) = K \Sigma^{1+1/n} \,,
\eqe
where $K$ and $n$ are positive constants, and $n$ is called the polytrope index. 
By normalization~\eref{eq:ball.normalization}, the mass density at center is unity, $\Sigma\ls{c} = 1$. 
Therefore, the coefficient $K$ is equal to the pressure at center in our normalization~\eref{eq:ball.normalization},
\eqb
 P\ls{c} = K \,(= P(\Sigma=1)\,) \,.
\eqe
\seqe
The form~\eqref{eq:ball.eos} is the simple polytrope whose index, $n$, is fixed at one value for all region inside the fluid ball. 
We are going to extend the simple form~\eqref{eq:ball.eos} to the piecewise polytrope in section~\ref{sec:ce}, but in this section we focus on the simple form~\eqref{eq:ball.eos}.

The surface of fluid ball is defined by the zero pressure, $P_\ast = 0$, where the suffix $\ast$ denotes the value at surface. 
Therefore, the mass density at surface is zero due to polytropic equation of states~\eref{eq:ball.eos}, $\Sigma_\ast = 0$. 
The normalized mass density takes values in the interval,
\eqb
\label{eq:ball.interval}
 (\Sigma_\ast =)\, 0 \le \Sigma \le 1 \, (= \Sigma\ls{c}) \,.
\eqe
Regarding $\Sigma$ as an independent variable, this finite interval of $\Sigma$ seems to be useful for numerical analysis. 
Hence, we regard all variables as functions of $\Sigma$,
\eqb
 R=R(\Sigma) \quad,\quad M=M(\Sigma) \quad,\quad \Phi=\Phi(\Sigma) \,,
\eqe
and $P(\Sigma)$ is already expressed as a function of $\Sigma$ in~\eref{eq:ball.eos}. 
The surface radius $R_\ast$ and total mass $M_\ast$ of polytropic ball are determined by
\eqb
 R_\ast = R(\Sigma=0) \quad,\quad
 M_\ast = M(\Sigma=0) \,.
\eqe

The sound speed $V$ in polytropic ball is given by
\eqb
\label{eq:ball.V}
 V^2 = \od{P(\Sigma)}{\Sigma} = P\ls{c} \Bigl(1+\frac{1}{n}\Bigr) \Sigma^{1/n} \,.
\eqe
This sound speed is normalized by light speed. 
Obviously, this $V$ decreases from center ($\Sigma\ls{c} = 1$) to surface ($\Sigma_\ast = 0$). 
The highest sound speed is given at the center,
\eqb
\label{eq:ball.Vc}
 V\ls{c}^2 = P\ls{c} \Bigl(1+\frac{1}{n}\Bigr) \,.
\eqe
In this paper, we assume a subluminal condition of sound speed,
\eqb
\label{eq:ball.subluminal}
 V\ls{c} \le 1 \,.
\eqe
Here it may be fair to note that a possibility of omitting the subluminal condition~\eqref{eq:ball.subluminal} has been discussed for the nuclear matters of neutron stars~\cite{ref:hartle.1978}. 
The omission of condition~\eqref{eq:ball.subluminal} may be possible if the nuclear matter possesses the properties so that the matter temperature is zero and the sound wave is dispersed and dumped (absorbed) by matters instantaneously. 
Even though such properties are good approximate ones of nuclear matters in neutron stars, those properties may not be exact and correct for all region inside neutron stars. 
Furthermore, our analysis is not restricted to neutron stars but is designed to include any matter described by polytropic equation of stats under some conditions which are reasonable from the viewpoint of general relativity. 
Hence, let us require the subluminal-sound-speed condition~\eqref{eq:ball.subluminal} which we regard as a rigorous general relativistic property of matters.

The outside region of polytropic ball, $R > R_\ast$, is described by the Schwarzschild geometry of mass $M_\ast$. 
The inside region $R \le R_\ast$ is determined by the Einstein equation and conservation law $T^{\mu\nu}_{\phantom{\mu\nu};\nu} = 0$, which are reduced to the Tolman-Oppenheimer-Volkoff (TOV) equations,
\seqb
\eqab
\label{eq:ball.tov-1}
 \od{M}{\Sigma} &=&
 4 \pi R^2 \Sigma \od{R}{\Sigma}
\\
\label{eq:ball.tov-2}
 \od{P}{\Sigma} &=& - \frac{(\Sigma + P)\,(M + 4 \pi R^3 P)}{R (R - 2 M)} \od{R}{\Sigma}
\\
\label{eq:ball.tov-3}
 \od{\Phi}{\Sigma} &=& - \frac{1}{\Sigma + P} \od{P}{\Sigma} \,.
\eqae
\seqe
Two functions $R(\Sigma)$ and $M(\Sigma)$ are obtained by solving \eref{eq:ball.tov-1} and \eref{eq:ball.tov-2} under the equation of states~\eref{eq:ball.eos}. 
Substituting those solutions into \eref{eq:ball.tov-3}, $\Phi(\Sigma)$ is obtained. 
Those solutions of TOV equations depend on two parameters, $V\ls{c}$ and $n$, for the simple polytrope~\eqref{eq:ball.eos}.

Under the set-up given above, our aim is to analyze the problem, whether or not the properties (A) and (B), which are described in section~\ref{sec:intro}, hold for polytropic ball under the subluminal-sound-speed condition~\eref{eq:ball.subluminal}. 
In following analyses, TOV equations~\eref{eq:ball.tov-1} and~\eref{eq:ball.tov-2} are solved numerically. 
A technical remark for numerical calculation is summarized in the appendix~\ref{app:tov}, which is applied to sections~\ref{sec:a}, \ref{sec:b} and~\ref{sec:ce}. 
All of our numerical analyses are performed with Mathematica ver.10.

\section{Problem A: Can a UCOP appear outside the simple polytropic ball?}
\label{sec:a}

The problem in this section is whether or not an inequality, $R_\ast < 3M_\ast$, holds for the simple polytropic ball of equation of states~\eqref{eq:ball.eos} under the condition~\eref{eq:ball.subluminal}. 
If $R_\ast < 3M_\ast$, then a UCOP appears outside a polytropic ball. 
Otherwise, if $3M_\ast < R_\ast$, then a UCOP does not appear outside the ball. 
Our strategy is as follows:
\begin{description}
\item[A1: ]
Solve numerically TOV equations~\eref{eq:ball.tov-1} and~\eref{eq:ball.tov-2} for given values of parameters $(V\ls{c},n)$, and calculate the mass-to-radius ratio, $3M_\ast/R_\ast\,$.
\item[A2: ]
Iterate the step A1 with varying parameters $(V\ls{c},n)$, so as to obtain the ratio, $3M_\ast/R_\ast$, as a function of parameters $(V\ls{c},n)$.
\item[A3: ]
Find the maximum value of $3M_\ast/R_\ast$ as a function of $(V\ls{c},n)$.
If the maximum is less than unity, we conclude that the inequality, $3M_\ast < R_\ast$, holds for all values of $(V\ls{c},n)$, and no UCOP appears outside the simple polytropic ball of equation of states~\eqref{eq:ball.eos} under the subluminal-sound-speed condition~\eqref{eq:ball.subluminal}.
\end{description}

In the Newton gravity, the total mass and surface radius of the simple polytropic ball are finite in the interval of index, $0 < n < 5$, but diverge in the interval, $5 \le n$, for any value of $V\ls{c} >0$. 
However, in the Einstein gravity, the thorough numerical analysis of simple polytropic ball by Nilsson and Uggla~\cite{ref:nilsson+1.2001} have revealed a complicated behavior of $M_\ast$ and $R_\ast$ in the half-infinite interval of central sound speed, $0 < V\ls{c}\,$:
\begin{itemize}
\item
In the interval of polytrope index, $0 < n < 3.339$, both of $M_\ast$ and $R_\ast$ are finite.
\item
In the interval, $3.339 \le n < 5$, a complicated behavior is found. 
\begin{itemize}
\item
Both of $M_\ast$ and $R_\ast$ are finite for almost of all values of $(V\ls{c},n)$ in the present parameter region. 
\item
However, both of $M_\ast$ and $R_\ast$ diverge at some discrete points $(V\ls{c}^{(i)},n^{(i)})$ in $V\ls{c}$-$n$ plane, where $i = 1, 2, \cdots, N\ls{div}$. 
The number of such divergence points, $N\ls{div}$ ($=$ finite or countable infinity), cannot be read from Nilsson-Uggla~\cite{ref:nilsson+1.2001}. 
\end{itemize}
Note that, although the existence of some divergence points $(V\ls{c}^{(i)},n^{(i)})$ in $V\ls{c}$-$n$ plane has been definitely confirmed, the accurate positions of them have not been specified.\footnote{
Two examples of such divergence points are $(V\ls{c},n) = (V\ls{c}^{(1)},3.357) \,,\, (V\ls{c}^{(2)},4.414)$, where the values $V\ls{c}^{(1)}$ and $V\ls{c}^{(2)}$ cannot be read from Nilsson-Uggla~\cite{ref:nilsson+1.2001}.}
\item
In the interval, $5 \le n$, both of $M_\ast$ and $R_\ast$ are infinity. 
\end{itemize}
Note that the mass-to-radius ratio has not been analyzed in Nilsson-Uggla~\cite{ref:nilsson+1.2001}.
The analysis of the ratio is our task.

\begin{figure}[t]
 \begin{center}
 \includegraphics[height=95mm]{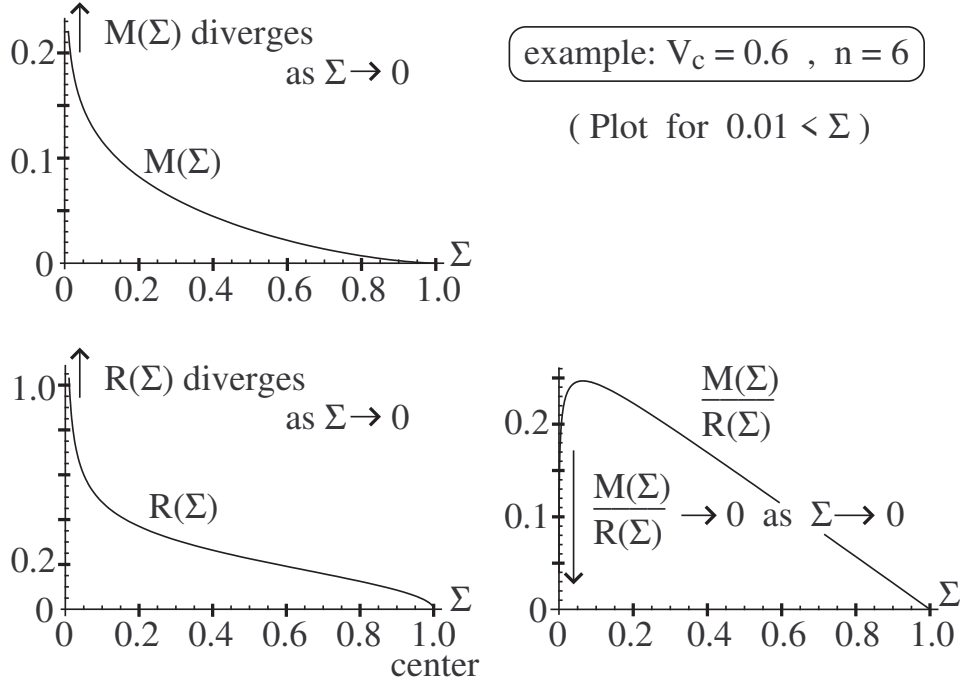}
 \end{center}
\caption{Numerical solution of TOV equations with the simple polytropic equation of states~\eqref{eq:ball.eos} at $(V\ls{c},n) = (0.6,6)$. 
Although $M_\ast$ and $R_\ast$ are infinity, the mass-to-radius ratio $M_\ast/R_\ast$ becomes zero.
}
\label{fig:n6}
\end{figure}

From the above behavior of $M_\ast$ and $R_\ast$ found by Nilsson-Uggla, the physically interesting region of parameters are
\eqb
\label{eq:a.region}
 0 < V\ls{c} \le 1 \quad,\quad 0 < n < 5 \,,
\eqe
where the interval of $V\ls{c}$ denotes the subluminal-sound-speed condition~\eref{eq:ball.subluminal}. 
It is enough for our aim to calculate the ratio, $3M_\ast/R_\ast$, in this parameter region. 
However, the parameter points $(V\ls{c}^{(i)},n^{(i)})$, where $M_\ast$ and $R_\ast$ diverge in the region~\eref{eq:a.region}, may not be included in the grid points of numerical analysis (see the step A2 of our strategy). 
In order to guess the behavior of $3M_\ast/R_\ast$ at those points $(V\ls{c}^{(i)},n^{(i)})$, we observe the solutions of TOV equations in the interval, $5 \le n$, where $M_\ast$ and $R_\ast$ diverge as well. 
\Fref{fig:n6} is an example with $V\ls{c}= 0.6$ and $n = 6$. 
This figure shows that, although the mass $M(\Sigma)$ and radius $R(\Sigma)$ diverge as the surface ($\Sigma = 0$) is approached, the mass-to-radius ratio $M(\Sigma)/R(\Sigma)$ converges to zero. 
The same behavior is observed for the other values of $(V\ls{c},n)$ in the interval, $5 \le n$. 
Hence, it is expected, even at the parameter points $(V\ls{c}^{(i)},n^{(i)})$ where $M_\ast$ and $R_\ast$ diverge in the region~\eref{eq:a.region}, the mass-to-radius ratio converges to zero.

\begin{figure}[t]
 \begin{center}
 \includegraphics[height=105mm]{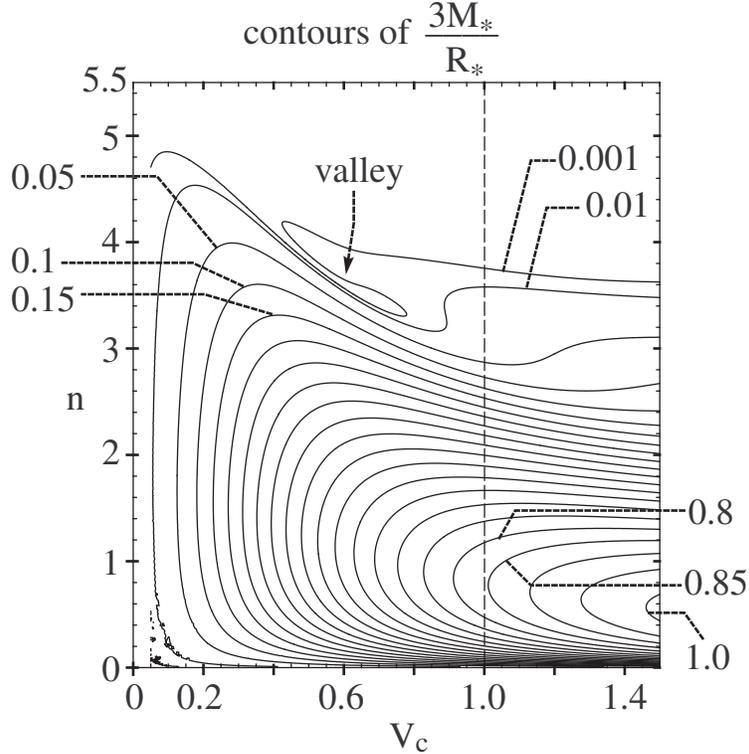}
 \end{center}
\caption{Contours of $3M_\ast/R_\ast$ on $V\ls{c}$-$n$ plane for the simple polytropic ball. 
Numerical calculation is performed in the region, $0.05 \le V\ls{c} \le 1.5$ and $0.01 \le n \le 5.5$. 
Obviously, the inequality, $3M_\ast/R_\ast < 1$, holds under the subluminal-sound-speed condition, $V\ls{c} \le 1$.
For small values of $V\ls{c}$ and $n$, numerical errors become manifest.}
\label{fig:3MtoR}
\end{figure}

With the help of above discussion, we can safely carry out our strategy of numerical analysis, composed of steps A1, A2 and A3. 
The result is shown in \fref{fig:3MtoR}, in which the contours of $3M_\ast/R_\ast$ are plotted. 
Although our main interest is in the parameter region~\eref{eq:a.region}, we have calculated for a bit larger region, $0.05 \le V\ls{c} \le 1.5$ and $0.01 \le n \le 5.5$. 
\footnote{
In \fref{fig:3MtoR}, the ratio $3M_\ast/R_\ast$ is small enough in the interval, $5 \le n \le 5.5$. 
This is consistent with the \fref{fig:n6} and discussion after \eref{eq:a.region}. 
Further, we have found numerical implications, although details are not shown here, that the points $(V\ls{c}^{(i)},n^{(i)})$, where $M_\ast$ and $R_\ast$ diverge in the region~\eref{eq:a.region}, form a line along the valley shown in \fref{fig:3MtoR}, and this line seems to approach $(V\ls{c},n) = (0,5)$.} 
It is obvious that the maximum value of $3M_\ast/R_\ast$ in the region~\eref{eq:a.region} appears on the vertical line at $V\ls{c} = 1$ in \fref{fig:3MtoR}. 
This maximum takes the value between $3M_\ast/R_\ast = 0.8$ and $0.85$. 
(A more precise value is calculated in section~\ref{sec:sd}.) 
Hence, we can conclude that the inequality, $3M_\ast/R_\ast < 1$, holds in the physically interesting region~\eref{eq:a.region}. 
No UCOP appears in the outside Schwarzschild geometry under the subluminal-sound-speed condition~\eref{eq:ball.subluminal} for the simple polytropic ball of equation of states~\eqref{eq:ball.eos}.

\section{Problem B: Can a UCOP appear inside the simple polytropic ball?}
\label{sec:b}

The problem in this section is whether a UCOP can exist inside the simple polytropic ball of equation of states~\eqref{eq:ball.eos} under the subluminal-sound-speed condition~\eref{eq:ball.subluminal}. 
For the first, the conditions for the existence of UCOP in static spherical spacetimes are summarized. 
Next, those conditions are applied to the simple polytropic ball, and we show that no UCOP can exit inside the simple polytropic ball.

\subsection{UCOP in static spherical spacetimes}

The feature of UCOP in the static spherical spacetime of metric~\eref{eq:ball.metric} is determined by the null geodesic equation. 
Denoting the affine parameter and radial coordinate of null geodesic by, respectively, $\lambda$ and $R\ls{null}(\lambda)$ under the normalization~\eref{eq:ball.normalization}, the radial component of null geodesic equation on spacetime~\eref{eq:ball.metric} is
\eqb
\label{eq:b.null}
 \Bigl(\od{R\ls{null}}{\lambda}\Bigr)^2 + \omega^2 U\ls{eff}(R\ls{null}) = 0\,,
\eqe
where $U\ls{eff}$ is the effective potential given by
\eqb
\label{eq:b.Ueff}
 U\ls{eff}(R) =
 \Bigl[\frac{b^2}{R^2} - \exp\bigl(-2\Phi(R)\,\bigr)\, \Bigr]
 \,\Bigl(1-\frac{2 M(R)}{R}\Bigr)
 \quad,\quad
 b = \frac{l}{\omega}
 \,,
\eqe
where $M(R)$ and $\Phi(R)$ are regarded as functions of $R$ given by solving TOV equations, $b$ is an impact parameter, and $l$ and $\omega$ are respectively the orbital angular momentum and the frequency of photon measured at infinity.

\begin{figure}[t]
 \begin{center}
 \includegraphics[height=45mm]{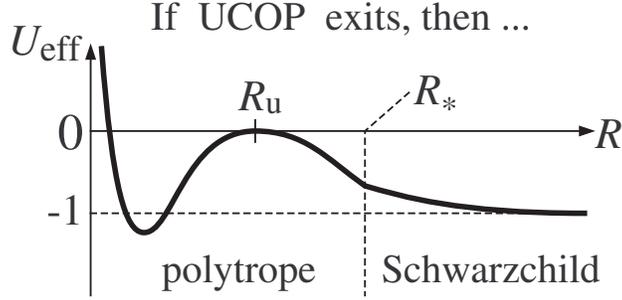}
 \end{center}
\caption{A schematic graph of $U\ls{eff}(R)$ if a UCOP exists inside the polytropic ball.}
\label{fig:Ueff}
\end{figure}

A photon propagating on a UCOP remains at a constant radius ($\diff R\ls{null}/\diff \lambda = 0$), but it is unstable. 
Hence, the radius of UCOP, $R\ls{u}$, is determined by following conditions,
\eqb
\label{eq:b.ucop.primitive}
 U\ls{eff}(R\ls{u}) = 0
 \quad,\quad
 \od{U\ls{eff}}{R}(R\ls{u}) = 0
 \quad,\quad
 \od{^2 U\ls{eff}}{R^2}(R\ls{u}) \le 0 \,.
\eqe
This implies that, if a UCOP exists inside the polytropic ball, the top of potential barrier touches below the zero level at $R\ls{u}$ as shown in \fref{fig:Ueff}.

By substituting \eref{eq:b.Ueff} into \eref{eq:b.ucop.primitive}, we obtain
\seqb
\eqab
\label{eq:b.ucop-1}
 && \frac{2}{R\ls{u}}F(R\ls{u}) - \od{F}{R}(R\ls{u}) = 0
\\
\label{eq:b.ucop-2}
 && \frac{1}{R\ls{u}}\od{F}{R}(R\ls{u}) - \od{^2F}{R^2}(R\ls{u}) \ge 0
\\
\label{eq:b.ucop-3}
 && b^2 = \frac{R\ls{u}^2}{F(R\ls{u})} \,,
\eqae
\seqe
where
\eqb
 F(R) \defeq \exp\bigl(\,2\Phi(R)\,\bigr) \,\,\,(=-g_{00}) \,.
\eqe
The radius of UCOP, $R\ls{u}$, is determined by~\eref{eq:b.ucop-1} and~\eref{eq:b.ucop-2}, and then the impact parameter of null geodesic circulating on UCOP forever is obtained by~\eref{eq:b.ucop-3}.
Therefore, the existence condition of UCOP consists of two parts; an algebraic equation~\eref{eq:b.ucop-1} and an inequality~\eref{eq:b.ucop-2}, which do not include the impact parameter.

\subsection{Non-existence of UCOP inside the simple polytropic ball}

In order to apply the existence conditions of UCOP~\eref{eq:b.ucop-1} and~\eref{eq:b.ucop-2} to simple polytropic balls, we need a concrete functional form of $\Phi(\Sigma) = (1/2)\ln F(\Sigma)$. 
Substituting the equation of states~\eref{eq:ball.eos} into a TOV equation~\eref{eq:ball.tov-3},
\eqb
 \od{\Phi(\Sigma)}{\Sigma}
 = -\Bigl(1+\frac{1}{n}\Bigr) \frac{P\ls{c} \Sigma^{1/n-1}}{1+P\ls{c}\Sigma^{1/n}}
 = -(n+1)\,\od{\ln(1 + P\ls{c} \Sigma^{1/n})}{\Sigma} \,.
\eqe
The integration constant for this equation is determined by the junction condition of metric at the surface of polytropic ball, $\exp(2\Phi_\ast) = 1 - 2M_\ast/R_\ast$. 
We obtain
\eqb
\label{eq:b.F}
 F(\Sigma) = e^{2\Phi(\Sigma)} =
 \frac{F_\ast}{(1+P\ls{c}\Sigma^{1/n})^{2(n+1)}} =
 F_\ast \Bigl( 1 + \frac{P(\Sigma)}{\Sigma} \Bigr)^{-2(n+1)} \,,
\eqe
where $F_\ast = 1-2M_\ast/R_\ast\,$. 
Regarding $\Sigma$ as a function of $R$, which is given by solving TOV equations~\eref{eq:ball.tov-1} and~\eref{eq:ball.tov-2}, we obtain $F(R)$ as a function of $R$.

From~\eref{eq:b.F}, first and second differentials of $F(R)$ are calculated,
\eqb
\label{eq:b.F.diff}
\begin{split}
 \od{F(R)}{R} &= 2\frac{M + 4\pi R^3 P}{R (R - 2M)}\,F(R) \quad (\,> 0\,)
\\
 \od{^2 F(R)}{R^2} &=
 \frac{1}{R}
 \Biggl[ \frac{4M - R}{R - 2M}
  + 4\pi R^3 \Bigl( \frac{\Sigma + P}{R - 2M} + \frac{\Sigma + 3P}{M + 4\pi R^3 P} \Bigr)
 \,\Biggr]\,\od{F(R)}{R} \,,
\end{split}
\eqe
where TOV equations~\eref{eq:ball.tov-1}, \eref{eq:ball.tov-2} and~\eref{eq:ball.tov-3} are used. 
Substituting these differentials into the existence conditions of UCOP~\eref{eq:b.ucop-1} and~\eref{eq:b.ucop-2}, we obtain
\eqb
\label{eq:b.ucop.poly}
 \left.C_1\right|_{R=R\ls{u}} = 0 \quad,\quad \left.C_2\right|_{R=R\ls{u}} \ge 0 \,,
\eqe
where
\eqb
\label{eq:b.C}
\begin{split}
 C_1 &= R - 4\pi R^3 P - 3M
 \\
 C_2 &= 1 -
 \Biggl[ \frac{4M - R}{R - 2M}
  + 4\pi R^3
    \Bigl( \frac{\Sigma + P}{R - 2M} + \frac{\Sigma + 3P}{M + 4\pi R^3 P} \Bigr) \Biggr] \,.
\end{split}
\eqe
If there does not exist $R\ls{u}$ which satisfies the conditions~\eref{eq:b.ucop.poly} for any value of parameters $(V\ls{c},n)$ in the physically interesting region~\eref{eq:a.region}, then it is concluded that no UCOP can appear inside the polytropic balls.

\begin{figure}[t]
 \begin{center}
 \includegraphics[height=115mm]{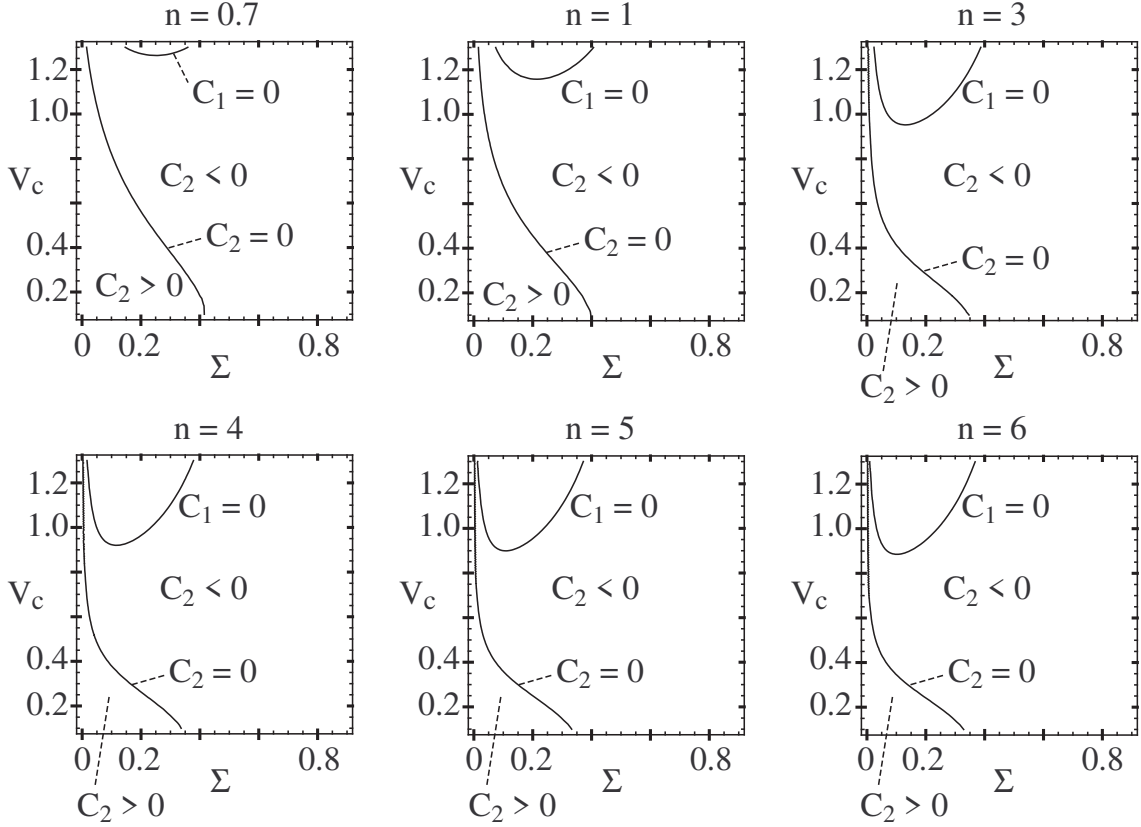}
 \end{center}
\caption{Plots of curves $C_1(\Sigma,V\ls{c},n) = 0$ and $C_2(\Sigma,V\ls{c},n) = 0$ in $\Sigma$-$V\ls{c}$ plane for some fixed $n$. 
The regions, $C_2 > 0$ and $C_2 < 0$, are also denoted.
It is recognized that the existence conditions of UCOP~\eref{eq:b.ucop.poly} are not satisfied for the single polytropic balls.
}
\label{fig:search}
\end{figure}

Note that the value of quantities $C_1$ and $C_2$ are calculated by substituting the solutions of TOV equations~\eref{eq:ball.tov-1} and~\eref{eq:ball.tov-2}. 
The solutions of TOV equations are functions of $\Sigma$ and depend on parameters $(V\ls{c},n)$. 
Hence, in our analysis, $C_1$ and $C_2$ can be obtained numerically as functions of three arguments, $C_1(\Sigma,V\ls{c},n)$ and $C_2(\Sigma,V\ls{c},n)$. 
Then, in order to check whether or not there exists $R\ls{u}$ satisfying~\eref{eq:b.ucop.poly}, our strategy is as follows:

\begin{description}
\item[B1: ]
Solve numerically TOV equations~\eref{eq:ball.tov-1} and~\eref{eq:ball.tov-2} for given value of parameters $(V\ls{c},n)$, and iterate this numerical calculation with varying $V\ls{c}$ and fixing $n$ at a given value. 
This iteration produces $C_1(\Sigma,V\ls{c},n)$ and $C_2(\Sigma,V\ls{c},n)$ as functions of $(\Sigma,V\ls{c})$ for the given value of $n$. 
\item[B2: ]
Plot two curves, $C_1 = 0$ and $C_2 = 0$, and identify two regions, $C_2 > 0$ and $C_2 <0$, in $\Sigma$-$V\ls{c}$ plane for the given $n$.
If the curve $C_1 = 0$ does not intersect with the region $C_2 \ge 0$, then it is concluded that no UCOP exists inside the simple polytropic ball at the given value of $n$. 
\item[B3: ]
Iterate the steps B1 and B2 with varying $n$, and check whether or not the intersection of $C_1 = 0$ with $C_2 \ge 0$ exists at each value of $n$. 
If the intersection does not appear for any value of $n$, then we conclude that a UCOP can never appear inside the simple polytropic ball of equation of states~\eqref{eq:ball.eos}.
\end{description}
The numerical result of this strategy is shown in \fref{fig:search}. 
We find that the curve $C_1 = 0$ remains in the region $C_2 < 0$ for all values of $n = 0.7 \,,\, 1 \,,\, 3 \,,\, 4 \,,\, 5 \,,\, 6$. 
The same feature is found for the other values of $n$ as far as we have calculated. 
Hence, no UCOP appears inside the polytropic ball. 
Furthermore, the non-existence of UCOP inside polytropic balls seems to hold for not only the physically interesting parameter region~\eref{eq:a.region} but also all parameter region, $0 < n$ and $0 < V\ls{c}$.

\section{Extension to the core-envelope piecewise polytrope models}
\label{sec:ce}

In this section, the analysis of simple polytrope model performed so far are extended to a representative model of the core-envelope piecewise polytropic fluid balls. 
For the first, the formulation of our model is described in section~\ref{sec:ce.model}. 
Then, in sections~\ref{sec:ce.a} and~\ref{sec:ce.b}, we investigate whether or not the properties (A) and (B), which are described in section~\ref{sec:intro}, hold for our model of the piecewise polytropic balls. 
Our analysis will indicate the same result with the simple polytropic ball that no UCOP appears inside and outside the representative model of the core-envelope piecewise polytropic balls.

\subsection{Our model of the core-envelope piecewise polytropic perfect fluid ball}
\label{sec:ce.model}

\begin{figure}[t]
 \begin{center}
 \includegraphics[height=45mm]{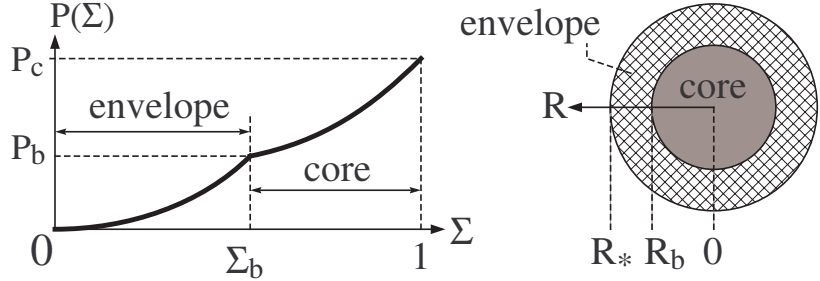}
 \end{center}
\caption{Left panel is a schematic graph of the core-envelope piecewise polytropic equation of states~\eqref{eq:ce.eos}. 
Right panel is a schematic image of the core-envelope model of fluid ball.}
\label{fig:CoreEnvelope}
\end{figure}

Dividing the interval of normalized mass density ($0 \le \Sigma \le 1$) into some sub-intervals, the piecewise polytropic equation of states is defined as the one whose polytrope index takes a different value at each sub-interval~\cite{ref:muller+1.1985}. 
The core-envelope type of it is the polytrope composed of two sub-intervals (see figure~\ref{fig:CoreEnvelope}),
\eqb
\label{eq:ce.eos}
 P(\Sigma) =
 \begin{cases}
  P\ls{c}\, \Sigma^{1+1/\ncor} & \text{in the core region, $\Sigma\ls{b} \le \Sigma \le 1$}
 \\
  \Kenv\, \Sigma^{1+1/\nenv} & \text{in the envelope region, $0 \le \Sigma < \Sigma\ls{b}$} \,,
 \end{cases}
\eqe
where $\Sigma\ls{b}$ is the mass density at the boundary between core and envelope regions, the parameters $\ncor$ and $\nenv$ are the polytrope indices of respectively the core and envelope regions, and $P\ls{c}$ is the pressure at center. 
We require that the pressure is continuous at the boundary between two regions, $P(\Sigma\ls{b} -0) = P(\Sigma\ls{b} +0)$, in order to maintain the mechanical balance at the boundary. 
By this junction condition, the coefficient $\Kenv$ is given by the other parameters,
\eqb
\label{eq:ce.Kenv}
 \Kenv = P\ls{c}\, \Sigma\ls{b}^{1/\ncor - 1/\nenv} \,.
\eqe
This implies that $\diff P/\diff\Sigma$ becomes discontinuous at $\Sigma\ls{b}$ as shown in figure~\ref{fig:CoreEnvelope}. 
And, through the TOV equation~\eqref{eq:ball.tov-2}, $\diff R/\diff\Sigma$ becomes discontinuous at $\Sigma\ls{b}$.

If we would restrict our discussion to neutron stars, the nuclear matter near the surface might be in a solid state, and the matter around the core might be in a fluid state~\cite{ref:hartle.1978}.\footnote{
When going down from the surface to the center of neutron star, there may occur a phase transition from a solid state to a fluid state at some radius $R\ls{b} = R(\Sigma\ls{b})$ due to the increase of matter density in the inward direction.} 
Such inner structure of neutron star may be approximately modeled by applying the above equation of states~\eqref{eq:ce.eos}, and the phase transition density $\Sigma\ls{b}$ is determined by some reasonable nuclear matter physics. 
However, our discussion is not only for neutron stars but also for general spherical static balls of equation of states~\eqref{eq:ce.eos} under the subluminal-sound-speed condition. 
Therefore, we regard $\Sigma\ls{b}$ as a free parameter in our analysis.

The sound speed $V$ is given by $V^2 = \diff P/\diff \Sigma$. 
Then, there are two local maximums of $V$ at the center ($\Sigma = 1$) and the boundary approached from the envelope ($\Sigma \to \Sigma\ls{b}-0$). 
Hence, the subluminal-sound-speed condition is given by
\eqb
\label{eq:ce.subluminal}
\begin{split}
 V\ls{c}^2 &\defeq P\ls{c} \Bigl(1+\frac{1}{\ncor}\Bigr) \le 1
\\
 V\ls{b}^2 &\defeq \Kenv \Bigl(1+\frac{1}{\nenv}\Bigr) \Sigma\ls{b}^{1/\nenv}
 = P\ls{c} \Bigl(1+\frac{1}{\nenv}\Bigr) \Sigma\ls{b}^{1/\ncor} \le 1 \,,
\end{split}
\eqe
where the relation~\eqref{eq:ce.Kenv} is used in the second equality for $V\ls{b}$.

Here let us note that, according to our previous paper~\cite{ref:fujisawa+3.2015} which analysed the mass-to-radius ratio $3M_\ast/R_\ast$ of generic perfect fluid balls of any equation of states under the subluminal-sound-speed condition, the ratio $3M_\ast/R_\ast$ tends to decrease as the sound speed increases inside the ball. 
Therefore, in search of the upper bound of $3M_\ast/R_\ast$ for the core-envelope piecewise polytropic balls, it seems to be reasonable to set the sound speed at $\Sigma = 1$ and $\Sigma\ls{b}$ being the light speed,
\eqb
\label{eq:ce.extreme}
 V\ls{c} = 1 \quad,\quad V\ls{b} = 1 \,.
\eqe
The polytropic ball under this extreme-sound-speed condition is the model we focus on in the following analyses.

The condition~\eqref{eq:ce.extreme} and the relation~\eqref{eq:ce.Kenv} give the relations,
\seqb
\label{eq:ce.ourmodel}
\eqab
 P\ls{c}(\ncor) &=&
 \frac{\ncor}{\ncor + 1}
\\
 \nenv(\Sigma\ls{b},\ncor) &=&
 \Bigl[ \frac{\ncor + 1}{\ncor} \Sigma\ls{b}^{-1/\ncor} -1 \Bigr]^{-1}
\\
 \Kenv(\Sigma\ls{b},\ncor) &=&
 \frac{\ncor}{\ncor + 1} \Sigma\ls{b}^{1/\ncor-1/\nenv} \,.
\eqae
\seqe
Through these relations, we find that the core-envelope piecewise polytrope~\eqref{eq:ce.eos} depends on a variable $\Sigma$ and two free parameters $(\Sigma\ls{b},\ncor)$. 
Actually, our equation of states~\eqref{eq:ce.eos} is the function of three arguments, $P(\Sigma,\Sigma\ls{b},\ncor)$, under the extreme-sound-speed condition~\eqref{eq:ce.extreme}.

The parameter $\Sigma\ls{b}$ takes, obviously, a value in the interval,
\seqb
\label{eq:ce.region}
\eqb
 0 < \Sigma\ls{b} < 1 \,.
\eqe
On the other hand, the physically interesting interval of the piecewise polytrope index, $\ncor$, is not obvious. 
Remember that, for the simple polytrope~\eqref{eq:ball.eos}, the physically interesting interval of polytrope index ($0<n<5$) has been determined by Nilsson-Uggla's thorough numerical analysis of the simple polytrope~\cite{ref:nilsson+1.2001}. 
Therefore, the analysis of Nilsson-Uggla should be extended to the piecewise polytropic balls when we need an accurate interval of the index $\ncor$. 
However, in the following analyses, we simply assume that the physically reasonable interval of piecewise polytrope index $\ncor$ is the same with the simple polytrope index,
\eqb
 0 < \ncor < 5 \,.
\eqe
\seqe
In this paper, we do not aim the thorough analysis of the core-envelope piecewise polytropic balls, but we perform a test analysis with the representative model under the extreme-sound-speed condition~\eqref{eq:ce.extreme} in the parameter region~\eqref{eq:ce.region}. 
We expect that our model can represent some typical behavior of core-envelope piecewise polytropic balls.

Furthermore, here we introduce one more expectation: 
Remember that, for the simple polytrope~\eqref{eq:ball.eos} as mentioned in section~\ref{sec:a}, Nilsson-Uggla~\cite{ref:nilsson+1.2001} have revealed the existence of divergence of the mass $M_\ast$ and radius $R_\ast$ at some isolated parameter points in the physically interesting parameter region~\eqref{eq:a.region}. 
Therefore, for the core-envelope piecewise polytrope~\eqref{eq:ce.eos}, the same divergence of $M_\ast$ and $R_\ast$ may occur. 
However, let us expect that the ratio $3M_\ast/R_\ast$ converges to zero even if $M_\ast$ and $R_\ast$ of the piecewise polytropic ball diverge, since $3M_\ast/R_\ast$ of simple polytropic ball is expected to converge to zero as shown in figure~\ref{fig:n6}.

\subsection{Problem A: Can a UCOP appear outside the core-envelope piecewise polytropic ball?}
\label{sec:ce.a}

The problem which we numerically analyze in this section is whether or not an inequality, $R_\ast < 3M_\ast$, holds for the fluid ball of equation of states~\eqref{eq:ce.eos} under the extreme-sound-speed condition~\eref{eq:ce.extreme}. 
Our strategy to calculate the ratio $3M_\ast/R_\ast$ is as follows:
\begin{description}
\item[A1': ]
Solve numerically TOV equations~\eref{eq:ball.tov-1} and~\eref{eq:ball.tov-2} for given values of parameters $(\Sigma\ls{b},\ncor)$, and calculate the ratio, $3M_\ast/R_\ast\,$.
\item[A2': ]
Iterate the step A1' with varying parameters $(\Sigma\ls{b},\ncor)$, so as to obtain the ratio, $3M_\ast/R_\ast$, as a function of parameters $(\Sigma\ls{b},\ncor)$.
\item[A3': ]
Find the maximum value of $3M_\ast/R_\ast$ as a function of $(\Sigma\ls{b},\ncor)$.
If the maximum is less than unity, we conclude that the inequality, $3M_\ast < R_\ast$, holds for all values of $(\Sigma\ls{b},\ncor)$ in the parameter region~\eqref{eq:ce.region}, and no UCOP appears outside the core-envelope piecewise polytropic ball under the extreme-sound-speed condition~\eqref{eq:ce.extreme}. 
\end{description}

\begin{figure}[t]
 \begin{center}
 \includegraphics[height=90mm]{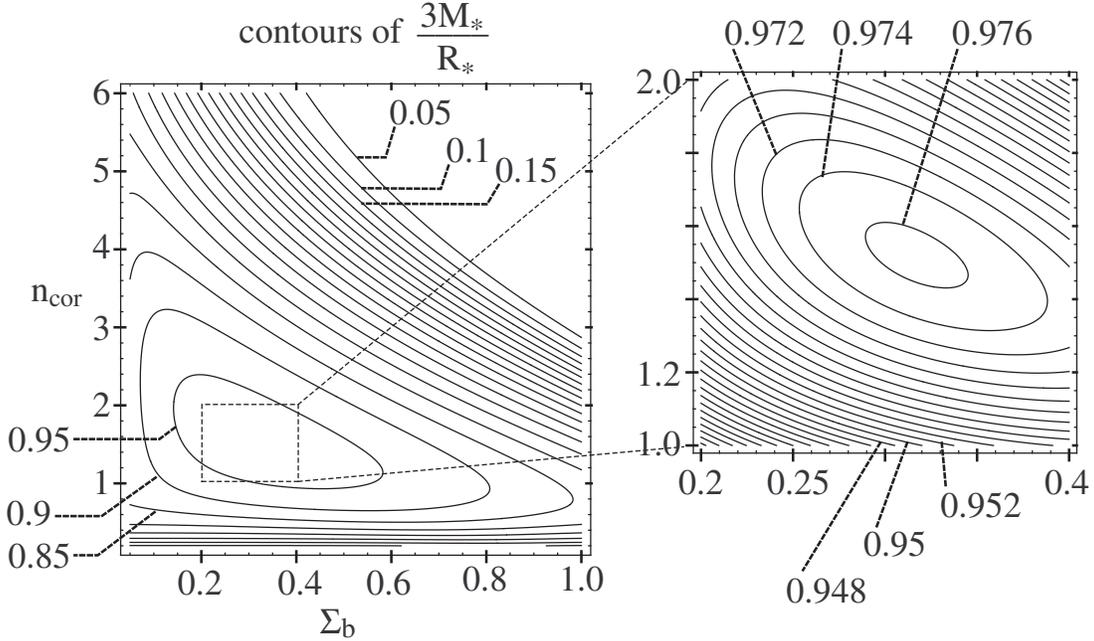}
 \end{center}
\caption{Contours of $3M_\ast/R_\ast$ on $\Sigma\ls{b}$-$\ncor$ plane for the core-envelope piecewise polytropic ball. 
Numerical calculation is performed in the region, $0.05 \le \Sigma\ls{b} \le 1.0$ and $0.2 \le n \le 6$. 
Contours of $3M_\ast/R_\ast$ are plotted at intervals of $0.05$ in left panel and at intervals of $0.002$ in right panel. 
It is found that the inequality, $3M_\ast/R_\ast < 1$, holds.}
\label{fig:3MtoR-ce}
\end{figure}

Our numerical result is shown in figure~\ref{fig:3MtoR-ce}, in which the contours of $3M_\ast/R_\ast$ are plotted on $\Sigma\ls{b}$-$\ncor$ plane. 
Because the contour of $3M_\ast/R_\ast = 1$ does not appear, the inequality, $3M_\ast < R_\ast$, holds for our model of core-envelope piecewise polytropic ball under the extreme-sound-speed condition~\eqref{eq:ce.extreme} in our interesting parameter region~\eqref{eq:ce.region}.

Although the numerical analysis shown in \fref{fig:3MtoR-ce} has been performed under the extreme-sound-speed condition~\eqref{eq:ce.extreme}, our original interest is the case under the subluminal-sound-speed condition~\eqref{eq:ce.subluminal}. 
In order to consider the subluminal case, let us remember the discussion for introducing the extreme-sound-speed condition~\eqref{eq:ce.extreme}. 
Then, according to our previous paper~\cite{ref:fujisawa+3.2015}, we can expect that no UCOP appears outside the core-envelope piecewise polytropic balls under the subluminal-sound-speed condition~\eqref{eq:ce.subluminal} in our interesting parameter region~\eqref{eq:ce.region}.

\subsection{Problem B: Can a UCOP appear inside the core-envelope piecewise polytropic ball?}
\label{sec:ce.b}

The problem which we numerically analyze in this section is whether or not a UCOP can exist inside the fluid ball of equation of states~\eqref{eq:ce.eos} under the extreme-sound-speed condition~\eref{eq:ce.extreme}. 
The calculation and discussion until equation~\eqref{eq:b.C} in section~\ref{sec:b} is applicable to the core-envelope piecewise polytropic ball. 
Therefore, our task is to check whether or not the existence conditions~\eqref{eq:b.ucop.poly} hold.

Note that, for the core-envelope piecewise polytropic ball under the extreme-sound-speed condition~\eqref{eq:ce.extreme}, the quantities $C_1$ and $C_2$ in equation~\eqref{eq:b.C} can be regarded as functions of three arguments, $C_1(\Sigma,\Sigma\ls{b},\ncor)$ and $C_2(\Sigma,\Sigma\ls{b},\ncor)$, since the values of $C_1$ and $C_2$ are calculated by substituting the solution of TOV equations~\eqref{eq:ball.tov-1} and~\eqref{eq:ball.tov-2}. 
Then, our strategy is as follows:
\begin{description}
\item[B1': ]
Solve numerically TOV equations~\eref{eq:ball.tov-1} and~\eref{eq:ball.tov-2} for given value of parameters $(\Sigma\ls{b},\ncor)$, and iterate this numerical calculation with varying $\Sigma\ls{b}$ and fixing $\ncor$ at a given value. 
This iteration produces $C_1(\Sigma,\Sigma\ls{b},\ncor)$ and $C_2(\Sigma,\Sigma\ls{b},\ncor)$ as functions of $(\Sigma,\Sigma\ls{b})$ for the given value of $\ncor$. 
\item[B2': ]
Plot two curves, $C_1 = 0$ and $C_2 = 0$, and identify two regions, $C_2 > 0$ and $C_2 <0$, in $\Sigma$-$\Sigma\ls{b}$ plane for the given $\ncor$.
If the curve $C_1 = 0$ does not intersect with the region $C_2 \ge 0$, it is concluded that, at the given value of $\ncor$, no UCOP exists inside the core-envelope piecewise polytropic ball under the extreme-sound-speed condition~\eqref{eq:ce.extreme}. 
\item[B3': ]
Iterate the steps B1' and B2' with varying $\ncor$, and check whether or not the intersection of $C_1 = 0$ with $C_2 \ge 0$ exists at each value of $\ncor$. 
If the intersection does not appear for any value of $\ncor$ in our interesting parameter region~\eqref{eq:ce.region}, then we conclude that a UCOP can never appear inside the core-envelope piecewise polytropic ball under the extreme-sound-speed condition~\eqref{eq:ce.extreme}.
\end{description}

\begin{figure}[t]
 \begin{center}
 \includegraphics[height=75mm]{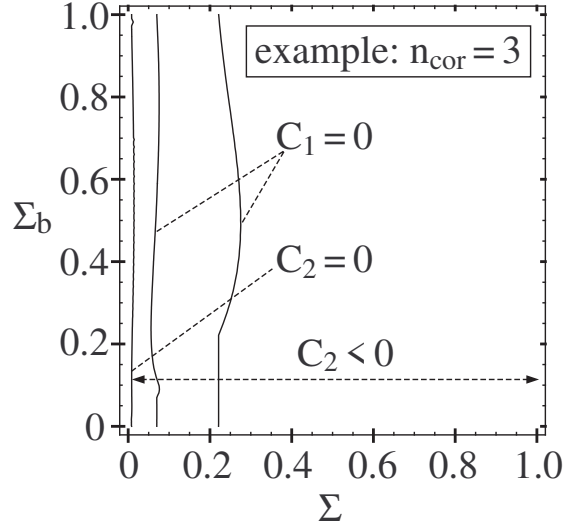}
 \end{center}
\caption{Plots of curves $C_1(\Sigma,\Sigma\ls{b},\ncor) = 0$ and $C_2(\Sigma,\Sigma\ls{b},\ncor) = 0$ in $\Sigma$-$\Sigma\ls{b}$ plane at polytrope index $\ncor = 3$. 
The region, $C_2 < 0$, is also denoted. 
It is recognized that the existence conditions of UCOP~\eref{eq:b.ucop.poly} are not satisfied for this case. 
}
\label{fig:search-ce}
\end{figure}

By the above strategy, it is found that the quantity $C_1(\Sigma,\Sigma\ls{b},\ncor)$ remains positive and does not become zero ($C_1 > 0$) in $\Sigma$-$\Sigma\ls{b}$ plane for low values of polytrope index, $0 < \ncor < n_{\text{cor-low}}$, where $n_{\text{cor-low}} \sim 1.5$. 
And, for higher values of polytrope index, $n_{\text{cor-low}} < \ncor < 5$, although the quantity $C_1$ becomes zero at some values of $\Sigma$ and $\Sigma\ls{b}$ in $\Sigma$-$\Sigma\ls{b}$ plane at every value of $\ncor$, the curve $C_1 = 0$ remains in the region $C_2 < 0$. 
The example at $\ncor = 3$ is shown in \fref{fig:search-ce}. 
Hence, we can conclude numerically that no UCOP appears inside the core-envelope piecewise polytropic balls under the extreme-sound-speed condition~\eqref{eq:ce.extreme} in our interesting parameter region~\eqref{eq:ce.region}.

Although the above numerical analysis has been performed under the extreme-sound-speed condition~\eqref{eq:ce.extreme}, our original interest is the case under the subluminal-sound-speed condition~\eqref{eq:ce.subluminal}. 
In order to consider the subluminal case, let us refer to the simple polytropic balls analysed in section~\ref{sec:b}, and note that the curve $C_1 = 0$ in $\Sigma$-$V\ls{c}$ plane for the simple polytrope, shown in figure~\ref{fig:search}, appears in the region of high values of $V\ls{c}$. 
Hence, if the similar behavior is expected for the core-envelope piecewise polytropic balls, the curve $C_1 = 0$ in $\Sigma$-$\Sigma\ls{b}$ plane tends to disappear as the sound speeds $V\ls{b}$ and $V\ls{c}$ decrease. 
That is, the intersection of $C_1 = 0$ with $C_2 < 0$ does not appear under not only the extreme-sound-speed condition~\eqref{eq:ce.extreme} but also the subluminal-sound-speed condition~\eqref{eq:ce.subluminal}. 
This discussion makes us expect that no UCOP appears inside the core-envelope piecewise polytropic balls under the subluminal-sound-speed condition~\eqref{eq:ce.subluminal} in our interesting parameter region~\eqref{eq:ce.region}.

\section{Summary and discussions}
\label{sec:sd}

We have investigated whether or not a UCOP can exist in the spacetime of static spherical ball of perfect fluid. 
The equations of states we have considered are the simple polytrope~\eqref{eq:ball.eos} and the core-envelope piecewise polytrope~\eqref{eq:ce.eos}. 
By numerical analyses of TOV equations and null geodesic equations, our result is as follows: 
For the simple polytropic balls, we have performed the thorough numerical investigation, and concluded that no UCOP can exist inside nor outside any simple polytropic ball under the subluminal-sound-speed condition. 
For the core-envelope piecewise polytropic balls, we have numerically investigated the representative model under the extreme-sound-speed condition~\eqref{eq:ce.extreme}, and concluded again that no UCOP can exist inside nor outside the core-envelope piecewise polytropic ball under the extreme-sound-speed condition. 
Further, according to our previous paper~\cite{ref:fujisawa+3.2015} and section~\ref{sec:b} of this paper, it is expected that the conclusion under the extreme-sound-speed condition is also true of the case under the subluminal-sound-speed condition.

Above conclusions mean that the polytropic balls investigated in this paper cannot be a black hole mimicker which possesses UCOP but no black hole horizon. 
This implies that, if the polytrope treated in this paper is a good model of the stellar matter in compact objects, the detection of shadow image by optical observation is regarded as a good evidence of black hole existence. 
Note that, to obtain a more definite conclusion for the core-envelope piecewise polytropic ball under the subluminal-sound-speed condition, a more detailed numerical research is necessary as discussed in the last two paragraphs of section~\ref{sec:ce.model}.

\begin{figure}[t]
 \begin{center}
 \includegraphics[height=50mm]{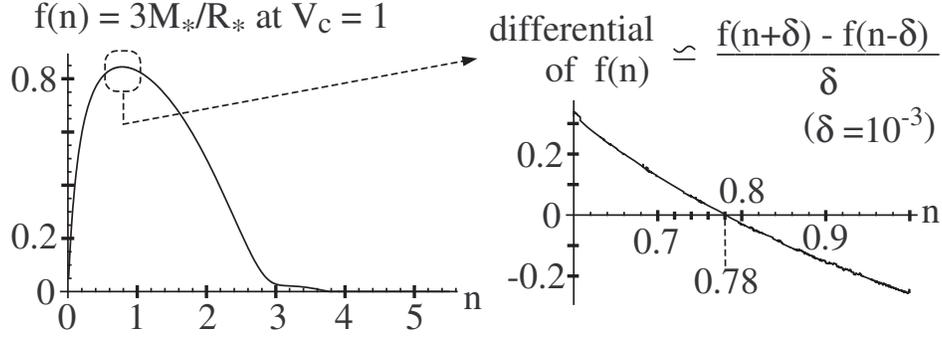}
 \end{center}
\caption{Sectioned diagram of \fref{fig:3MtoR} at $V\ls{c} = 1$. 
$f(n)$ describes $3M_\ast/R_\ast$ of the simple polytropic ball as a function of $n$ at $V\ls{c} = 1$. 
Differential of $f(n)$ is plotted by an approximation, $\diff f(n)/\diff n \simeq [f(n+\delta) - f(n-\delta)]/\delta$, where $\delta = 10^{-3}$. 
}
\label{fig:Vc1}
\end{figure}

Next, let us discuss a by-product of our analysis. 
In sections~\ref{sec:a} and~\ref{sec:ce.a}, the ratio of total mass to surface radius of polytropic ball, $3M_\ast/R_\ast$, has been the central issue. 
As mentioned in section~\ref{sec:intro}, the mass-to-radius ratio must be bounded above, $3M_\ast/R_\ast < 3/2$, in order to avoid gravitational collapse. 
Buchdahl~\cite{ref:buchdahl.1959} decreased the upper bound to, $3M_\ast/R_\ast < 4/3$, with assuming non-increasing mass density in outward direction and barotropic equation of states. 
Next, Barraco and Hamity~\cite{ref:barraco+1.2002} decreased the Buchdahl's upper bound to, $3M_\ast/R_\ast < 9/8$, by adding dominant energy condition to Buchdahl's assumptions. 
Furthermore, in our previous paper~\cite{ref:fujisawa+3.2015}, we decreased the Barraco-Hamity's upper bound to, $3M_\ast/R_\ast < 1.0909209$, by replacing the dominant energy condition with the subluminal-sound-speed condition. 
All these upper bounds remained greater than unity, which permits the existence of some black hole mimicker. 
However, as shown in \fref{fig:3MtoR} of this paper, the upper bound of $3M_\ast/R_\ast$ is decreased to a value lower than unity, by restricting the equation of states to the simple polytrope~\eref{eq:ball.eos} and assuming the subluminal-sound-speed condition~\eqref{eq:ball.subluminal}. 
Since the upper bound is found on the vertical line at $V\ls{c} = 1$ in \fref{fig:3MtoR}, a sectioned diagram of \fref{fig:3MtoR} at $V\ls{c} = 1$ is useful to read a precise value of the upper bound. 
It is shown in \fref{fig:Vc1}, where we define $f(n)$ by $3M_\ast/R_\ast$ as a function of $n$ at $V\ls{c} = 1$. 
From the \fref{fig:Vc1}, it is concluded that the following inequality holds in the physically interesting parameter region~\eref{eq:a.region} for the simple polytropic balls under the subluminal-sound-speed condition,
\eqb
\label{eq:sd.bound}
 \frac{3 M_\ast}{R_\ast} < 0.844 \,,
\eqe
where the upper bound is given by parameters $V\ls{c} = 1$ and $n \simeq 0.78$, and the value of upper bound is numerically calculated, $f(0.78) \simeq 0.844$. 
On the other hand, when we restrict the equation of states to the core-envelope piecewise polytrope~\eref{eq:ce.eos} and assuming the extreme-sound-speed condition~\eqref{eq:ce.extreme}, the upper bound of the ratio $3M_\ast/R_\ast$ can be roughly read from the \fref{fig:3MtoR-ce},
\eqb
\label{eq:sd.bound-ce}
 \frac{3 M_\ast}{R_\ast} < 0.976 \,.
\eqe

Finally, let us make a comment on a related topic. 
A possibility of trapping gravitational waves inside stellar objects has been discussed~\cite{ref:rosquist.1999,ref:andrade.2001}. 
The potential of gravitational perturbation is analysed in those discussions, while the potential of light propagation such as shown in \fref{fig:Ueff} is discussed in our analysis. 
So, when one is interested in a combination of optical observation and gravitational wave observation in the search of black holes and gravitational waves, the existence/non-existence conditions of UCOP and gravitational-wave-trapping may become an interesting issue.

\section*{Acknowledgements}

H.S. was supported by Japan Society for the Promotion of Science (JSPS), Grant-in-Aid for Scientific Research (KAKENHI, Exploratory Research, 26610050). 
Y.N. was supported in part by Japan Society for the Promotion of Science (JSPS), Grant-in-Aid for Scientific Research (KAKENHI, Grant no.15K05073). 
We would like to express our gratitude to the referee of this paper for a useful comment on the core-envelope type polytrope.

\appendix
\section{On numerical treatment of TOV equations}
\label{app:tov}

Right-hand sides of TOV equations~\eref{eq:ball.tov-1} and~\eref{eq:ball.tov-2} are indeterminate form at center because of the conditions, $M \to 0$ and $R \to 0$ as $\Sigma \to 1$. 
Therefore, in solving TOV equations numerically, we have made use of perturbative solutions near the center.

In order to consider a perturbation near center, we regard the radius $R$ as an independent variable, and the mass density as a function of radius, $\Sigma(R)$. 
TOV equations~\eref{eq:ball.tov-1} and~\eref{eq:ball.tov-2} are rearranged to
\eqb
\label{eq:tov.tov}
\begin{split}
 \od{M(R)}{R}
 &=
 4 \pi R^2 \Sigma(R)
\\
 \od{P(R)}{R}
 &=
 - \frac{\bigl[\,\Sigma(R) + P(R)\,\bigr]\,\bigl[\,M(R) + 4 \pi R^3 P(R)\,\bigr]}
        {R \,\bigl[\,R - 2 M(R)\,\bigr]} \,.
\end{split}
\eqe
For a sufficiently small radius $R \ll 1$, we introduce perturbations,
\eqb
\label{eq:tov.perturbation.primitive}
\begin{split}
 M(R) &= M\ls{(1)} R + M\ls{(2)} R^2 + M\ls{(3)} R^3 + \cdots
\\
 P(R) &= P\ls{c} + P\ls{(1)} R + P\ls{(2)} R^2 + P\ls{(3)} R^3 + \cdots
\\
 \Sigma(R) &= 1 + \Sigma\ls{(1)} R + \Sigma\ls{(2)} R^2 + \Sigma\ls{(3)} R^3 + \cdots \,,
\end{split}
\eqe
where conditions $M(R=0) = 0$, $\Sigma(R=0) = 1$ and $P(R=0) = P\ls{c}$ are included. 
Substituting~\eref{eq:tov.perturbation.primitive} into \eref{eq:tov.tov}, we obtain $M\ls{(1)} = 0$, $M\ls{2} = 0$, $P\ls{(1)} = 0$ and remaining parts,
\eqb
\label{eq:tov.perturbation}
\begin{split}
 M(R) &= \frac{4}{3}\pi R^3 + \pi \Sigma\ls{(1)} R^4 + \cdots
\\
 P(R)
 &= P\ls{c} - \frac{2}{3}\pi (1 + 3 P\ls{c})\,(1 + P\ls{c}) R^2
   - \frac{\pi}{9} (7 + 15 P\ls{c}) \Sigma\ls{(1)} R^3 + \cdots
\\
 \Sigma(R) &= 1 + \Sigma\ls{(1)} R + \Sigma\ls{(2)} R^2 + \Sigma\ls{(3)} R^3 + \cdots \,,
\end{split}
\eqe
where the central pressure $P\ls{c}$ and coefficients $\Sigma_{(n)}$ $(n = 1, 2, 3, \cdots)$ are determined by concrete form of equation of states.

Substitute these perturbative expansions into the equation of states~\eref{eq:ball.eos}, we obtain
\eqb
 \Sigma\ls{(1)} = 0
 \quad,\quad
 \Sigma\ls{(2)} = - \frac{n}{n+1}\,\frac{2 \pi (1+P\ls{c})\,(1+3 P\ls{c})}{3 P\ls{c}} \,.
\eqe
Hence, denoting a small radius by $R_\delta \ll 1$, the mass density $\Sigma_\delta$ and mass $M_\delta$ at $R=R_\delta$ are approximately given by $\Sigma_\delta = 1 + \Sigma\ls{(2)} R_\delta^2$ and $M_\delta = (4\pi/3) R_\delta^3\,$. 
If the mass density near center $\Sigma_\delta$ are given, then the others are determined by
\eqb
\label{eq:tov.IC}
 R_\delta = \sqrt{\frac{1-\Sigma_\delta}{|\Sigma\ls{(2)}|}}
 \quad,\quad
 M_\delta =
 \frac{4}{3}\pi \Bigl(\frac{1-\Sigma_\delta}{|\Sigma\ls{(2)}|}\Bigr)^{3/2}
 \,.
\eqe
In numerical calculation, we have solved TOV equations~\eref{eq:ball.tov-1} and~\eref{eq:ball.tov-2} for interval $0 < \Sigma \le \Sigma_\delta$ with initial condition~\eref{eq:tov.IC}. 
Also, we have checked the convergence of numerical solutions with varying $\Sigma_\delta$. 
All results in this paper are obtained using $\Sigma_\delta = 1 - 10^{-4}$.


\end{document}